\documentclass[
    ,final            
  ]
  {aipproc}

\layoutstyle{8x11double}

\begin{document}

\title{Gamma rays from star-forming regions}


\author{Gustavo E. Romero}{
  address={Instituto Argentino de Radioastronom{\'\i}a (CCT La Plata, CONICET),
C.C.5, 1894 Villa Elisa,  Buenos Aires, Argentina}
,altaddress={Facultad de Ciencias Astron\'omicas y Geof{\'\i}sicas, Universidad Nacional de La Plata, Paseo del Bosque, B1900FWA La Plata, Argentina} 
}

\begin{abstract}
Star-forming regions have been tentatively associated with gamma-ray sources since the early days 
of the COS B satellite. After the Compton Gamma-Ray Observatory, the statistical evidence for such 
an association has became overwhelming. Recent results from Cherenkov telescopes indicate that 
some  high-energy sources are produced in regions of active star formation like Cygnus OB2 and 
Westerlund 2. In this paper I will briefly review what kind of stellar objects can produce gamma-ray emission 
in star-forming regions and I will suggest that the formation process of massive stars could in principle
result in the production of observable gamma rays. 

\end{abstract}

\maketitle


\section{Introduction}

The identification of star-forming regions (SFRs) with gamma-ray sources has a rather long story that can be traced till the first attempts to make statistical correlation analysis between COS-B data on discrete gamma-ray sources and SFRs (e.g. \cite{Montmerle1979}). With the advent of the Compton Gamma-Ray Observatory and the Energetic Gamma-Ray Experiment Telescope (EGRET) the statistical evidence for a physical association became overwhelming (\cite{Kaaret1996,Romero1999}). In the early 1980s, theories of cosmic-ray acceleration in the strong winds of early-type stars were worked out in great detail and massive stars were proposed as potential gamma-ray sources (e.g. \cite{Casse1980,Voelk1982,White1985,Pollock1987}). However, other types of sources could contribute to the gamma-ray emission detected in the direction of SFRs, sources such as accreting black holes, pulsars, supernova remnants, and pulsar wind nebulae. It was not until the discovery of the extended unidentified source TeV J2032+4130 in Cygnus OB2 by the HEGRA array of imaging Cherenkov telescopes in 2002 \cite{Aharonian2002} that the hypothesis of a high-energy source powered by stellar effects in a SFR was considered as sound from an observational point of view. Since then, other two high-energy sources have been detected in SFRs: Westerlund 2 \cite{Aharonian2007} and W43 \cite{Chavez2008}. All this points out in the direction that early-type stars and the collective effects of their winds can accelerate particles beyond TeV energies and produce significant gamma-ray emission. In this review, I will discuss how gamma rays can be produced by stars in the different evolutionary stages of a SFR. But before, some account on the current ideas of star formation is in order.     

\section{The formation of massive stars}

\begin{figure}
  \includegraphics[height=.3\textheight]{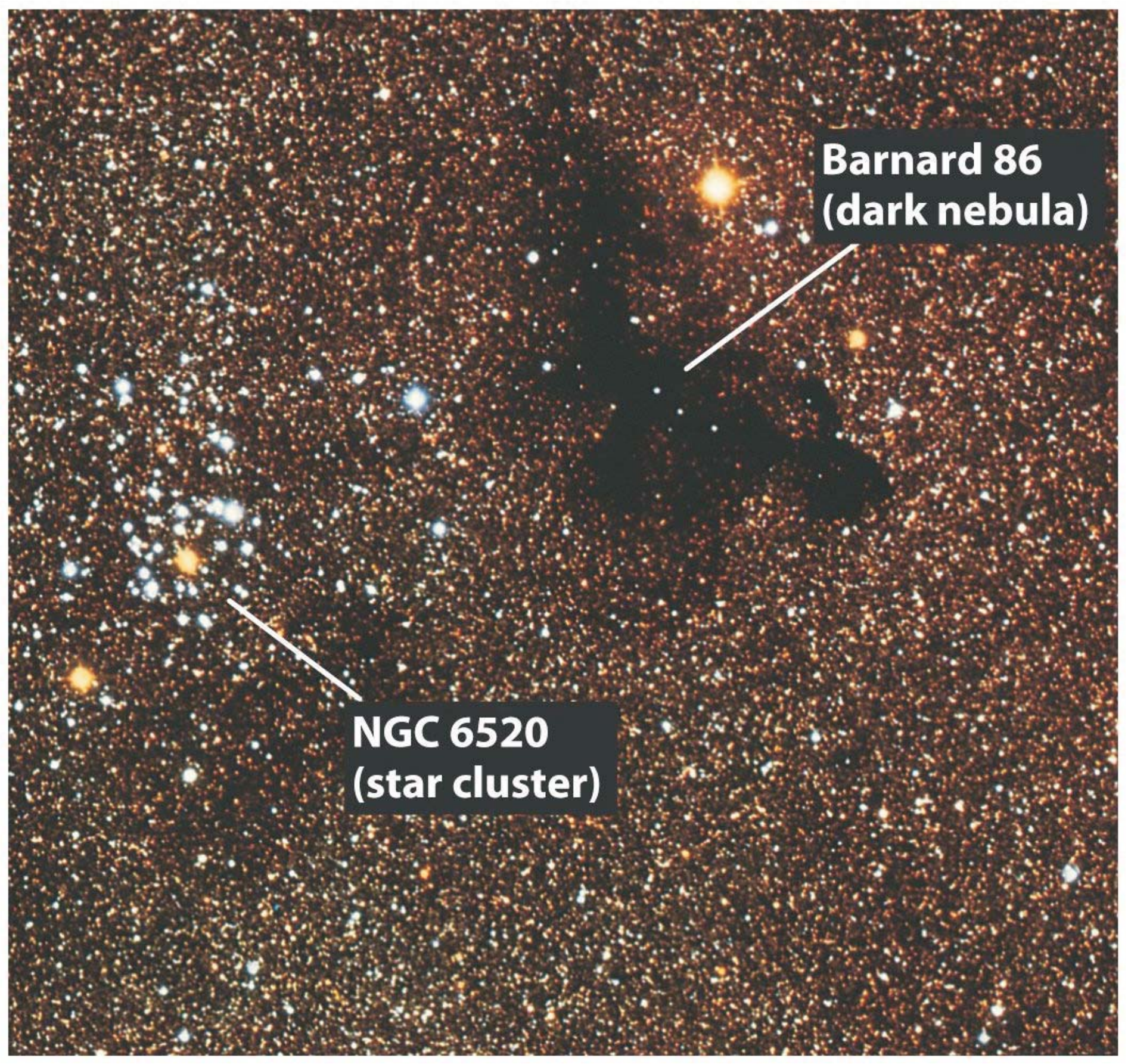}\label{fig1}
  \caption{Two different stages of a SFR evolution: a dark cloud and a fully formed open cluster.}
\end{figure}

The formation and early evolution of massive stars are not yet well understood. Two different mechanisms have been proposed. These mechanisms do not necessarily exclude each other. They are the accretion \cite{Osorio1999,McKee2003} and the coalescence \cite{Bonnell1998} massive star formation processes. In the
 first case the massive star is formed by accretion of gas in a dense molecular core. Such an accretion episode is always accompanied by ejection of thermal jets that can propagate inside the molecular cloud or even break out from it (for a review see, e.g., \cite{Garay1999}). In the coalescence picture the high-mass stars are formed by merging of low and intermediate mass protostars in the dense cluster environment. In either case the formation takes place inside a dense and giant molecular cloud. The SFR, then, starts as a giant dark cloud that is soon heated and illuminated by the star formation that goes on in its interior. As time goes by, the massive stars are formed and their strong winds sweep the ambient material giving rise to a mature open cluster. Both stages (initial and final) are illustrated in Fig. \ref{fig1}.

The dense molecular clouds that are the cradles of massive stars have relative small sizes ($<0.1$ pc), high temperatures that make of them strong IR sources ($T>100$ K), densities of the order of $\sim 10^7$ cm$^{-3}$ or more, and high luminosities in the range of $10^{4}-10^{6}$ $L_{\odot}$. The conditions of the medium are surely not homogeneous, with a power-aw density distribution $n\propto r^{-p}$, with $p$ in the range between 1.3 and 1.8 \cite{Garay2006}. The massive stars are formed under very high infall mass rates of the order of $10^{-3}-10^{-2}$ $M_{\odot}$ yr$^{-1}$. The whole formation process is rather rapid, with timescales of the order of a few kyr. 

One way to test the different scenarios proposed for massive star formation is the detection of outflows related to the accretion process in massive protostars. As we will see, there are some prominent examples of such outflows.       

\section{Collective effects of stellar winds in star forming regions }

Once a massive stellar association is formed, the combined effects of the stellar winds are expected to result in particle acceleration up to relativistic energies \cite{Bykov1,Bykov2,Bednarek2007}. In addition, supernova explosions of very massive (and hence quickly evolving) stars can help to create super-bubbles and collective shocks where particles can be efficiently accelerated (e.g. \cite{Parizot}). Hence, the average cosmic ray (CR) density in SFRs is expected to be higher than the galactic average. The locally generated CRs can interact with both extended passive targets, like molecular clouds or the rich intracluster medium (e.g. \cite{Fazio,Aharonian1996,Gabici}), or with the very same winds of the massive stars \cite{Torres2004}. In the latter case, a modulation effect due to convection is expected, and hence the low-energy emission would be greatly diminished or suppressed. Such a picture might correspond to that of extended TeV sources without significant correlated emission at low energies like TeV J2032+4130 \cite{Butt1,Butt2,Horns}. The same mechanism might operate in starburst galaxies where combined effect of the large number massive stars should produce a high density of CRs in a very rich environment \cite{Anchordoqui1999,Romero2003}.

\section{Gamma rays from massive stars and massive binary systems}

Individual hot stars with strong winds have been suggested also as potential sources of high energy particles and gamma rays (e.g. \cite{Casse1980,Voelk1982,White1985,White1991,White1992}. In some of these works particle acceleration occurs mainly in the end terminal shock and in others in shocks embedded in the winds. Some of these works predicted detections of isolated massive stars by EGRET, something that did not happen. Efficient electron and proton acceleration in the inner wind are unlikely, because of the strong losses to which both species are exposed during the processes. In addition, even if some gamma rays are produced they will be absorbed by the stellar photosphere producing electron-positron pairs. 

\begin{figure}
  \includegraphics[height=.3\textheight]{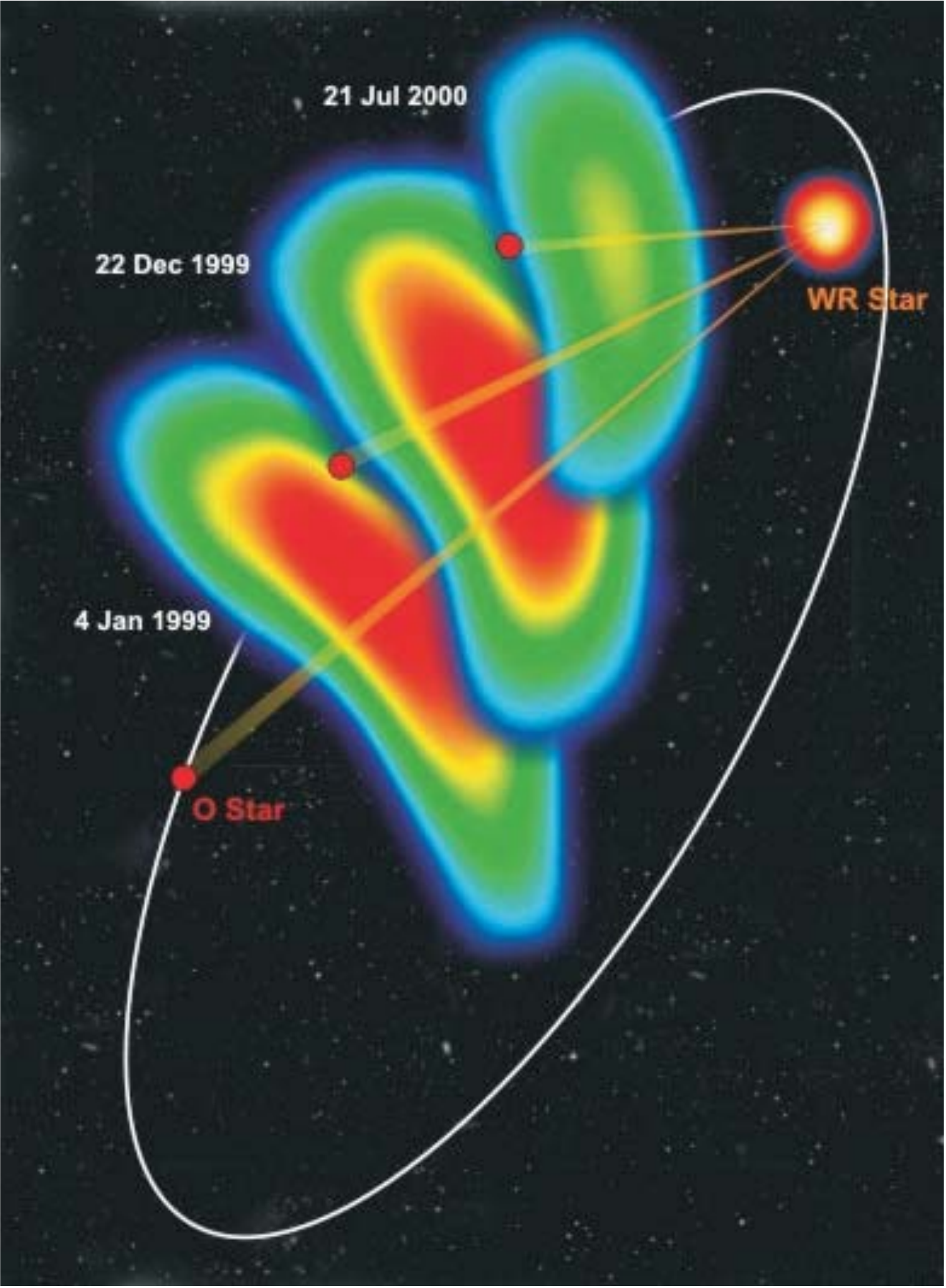}\label{fig2}
  \caption{Evolution of the colliding wind region at radio wavelength in WR140 \cite{Dougherty2005}.}
\end{figure}

Models based on colliding winds in massive binary systems, on the contrary, seem to be far more promising \cite{Eichler1993}. The clear detection of non-thermal radio emission from the colliding wind region in some systems like WR140, WR146, WR 147, and Cyg OB2 \#5 indicate that electrons are been accelerated up to relativistic energies in these binaries. These electrons are located relatively close to the secondary star, so they should have strong inverse Compton losses which might result in a detectable gamma-ray signal \cite{Benaglia2001,Benaglia2003}. Protons can be in principle also accelerate in the colliding wind shock and they can interact either with the wind material \cite{Pittard2006} or they can diffuse to interact with nearby molecular clouds \cite{Benaglia2005}. Detailed models have been recently developed for systems like WR140 \cite{Pittard2006,Reimer2006}, on the basis of the recent radio information obtained for this source \cite{Dougherty2005} (see Fig. \ref{fig2}), and for WR20a \cite{Bednarek2005}. This latter system, however, is not resolved at radio wavelengths and since it is very compact, strong high-energy photon absorption should take place.          

It is widely expected that GLAST-Fermi satellite and even perhaps AGILE could detect some colliding wind binaries. Depending on the orbital parameters, these sources might appear as highly variable at MeV-GeV energies. It is however dubious whether they can accelerate particles up to TeV energies.

\section{Gamma rays from massive protostars}

\begin{figure}
  \includegraphics[height=.5\textheight]{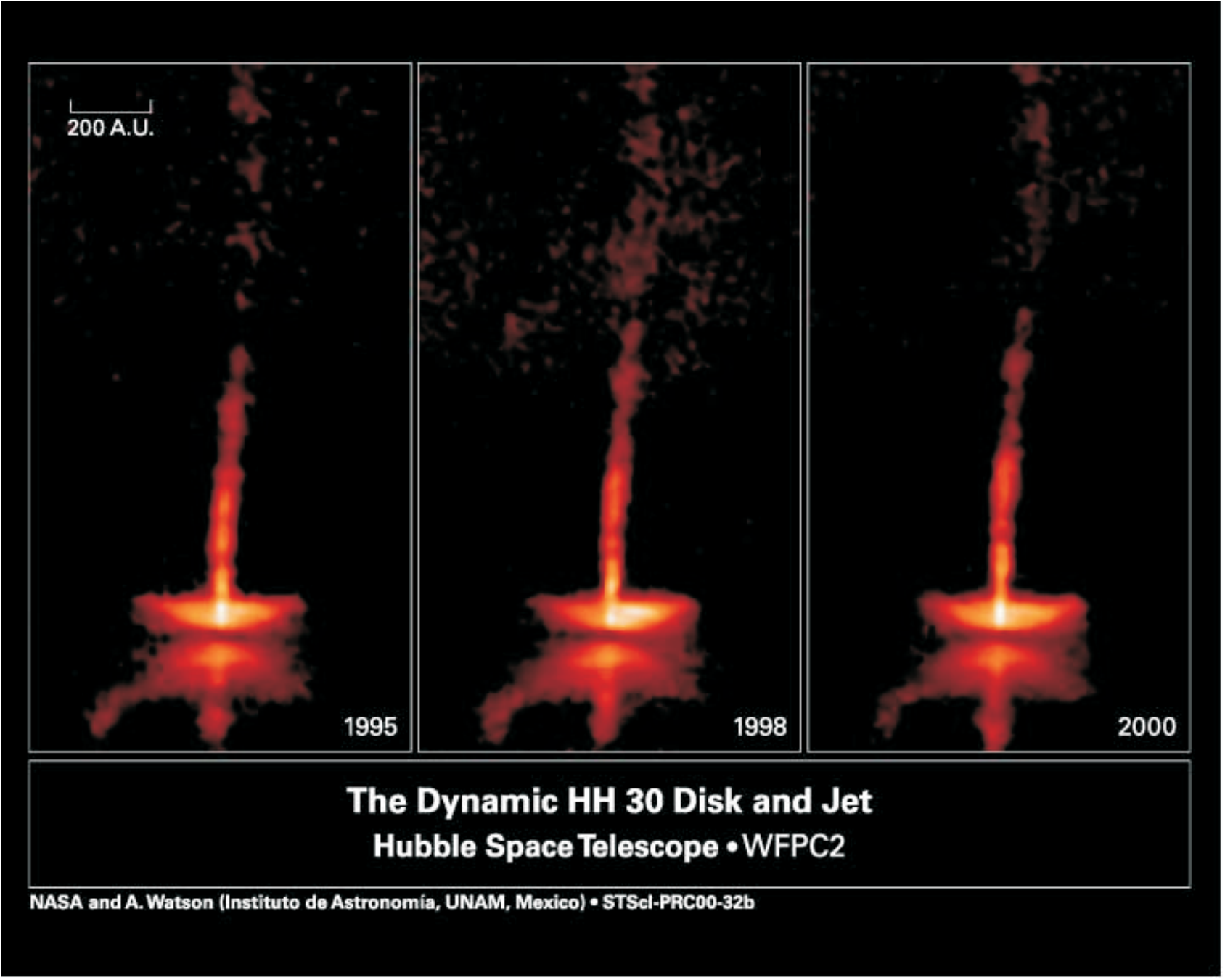}\label{fig3a}
  \caption{A protostellar disk and the associated jet as observed by the Hubble Space Telescope. Courtesy: NASA.}
\end{figure}

\begin{figure}
  \includegraphics[height=.4\textheight]{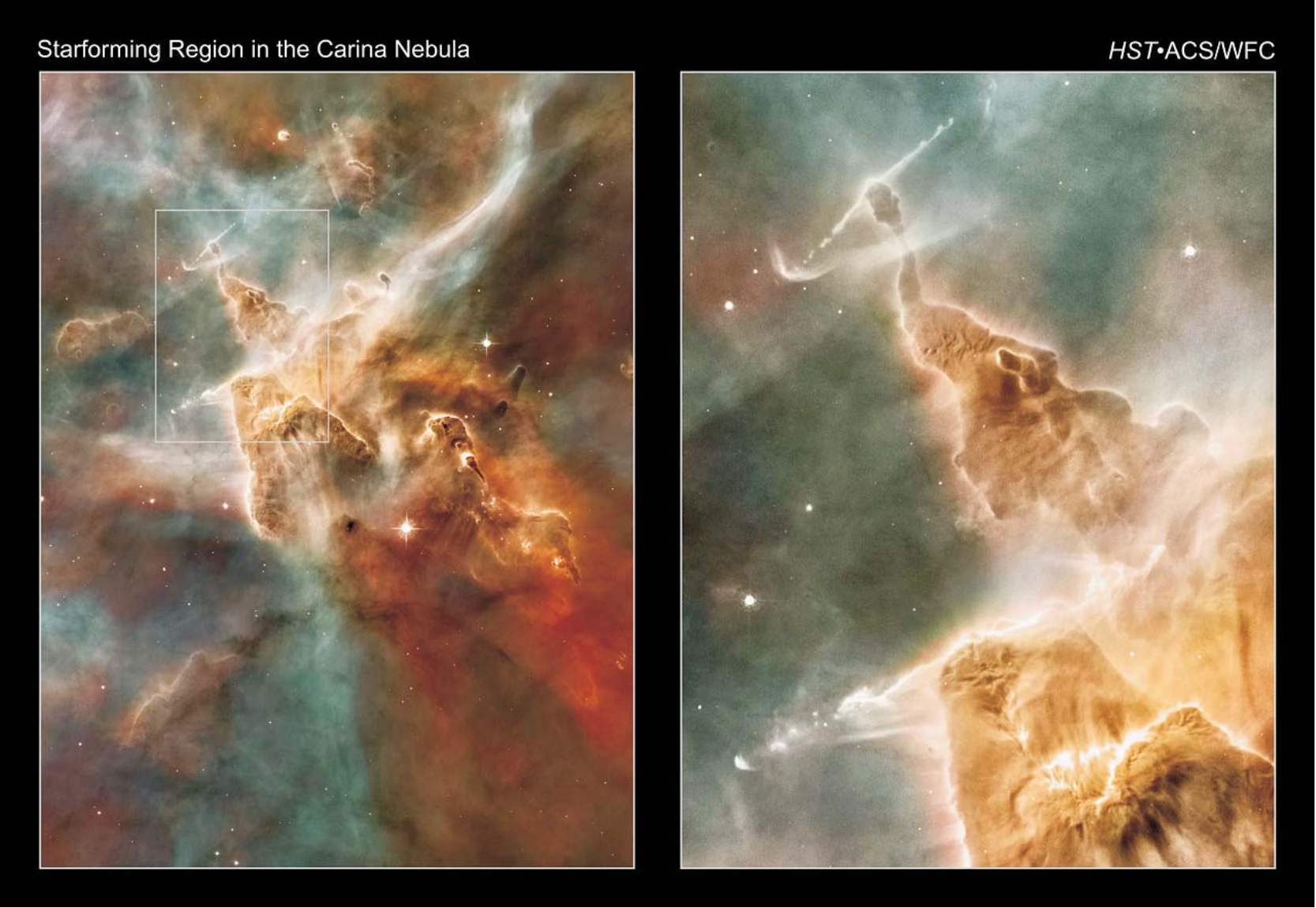}\label{fig3b}
  \caption{Star forming region in Carina, with several molecular cores and associated outflows. Image from the HST. Courtesy: NASA.}
\end{figure}

\begin{figure}
  \includegraphics[height=.4\textheight]{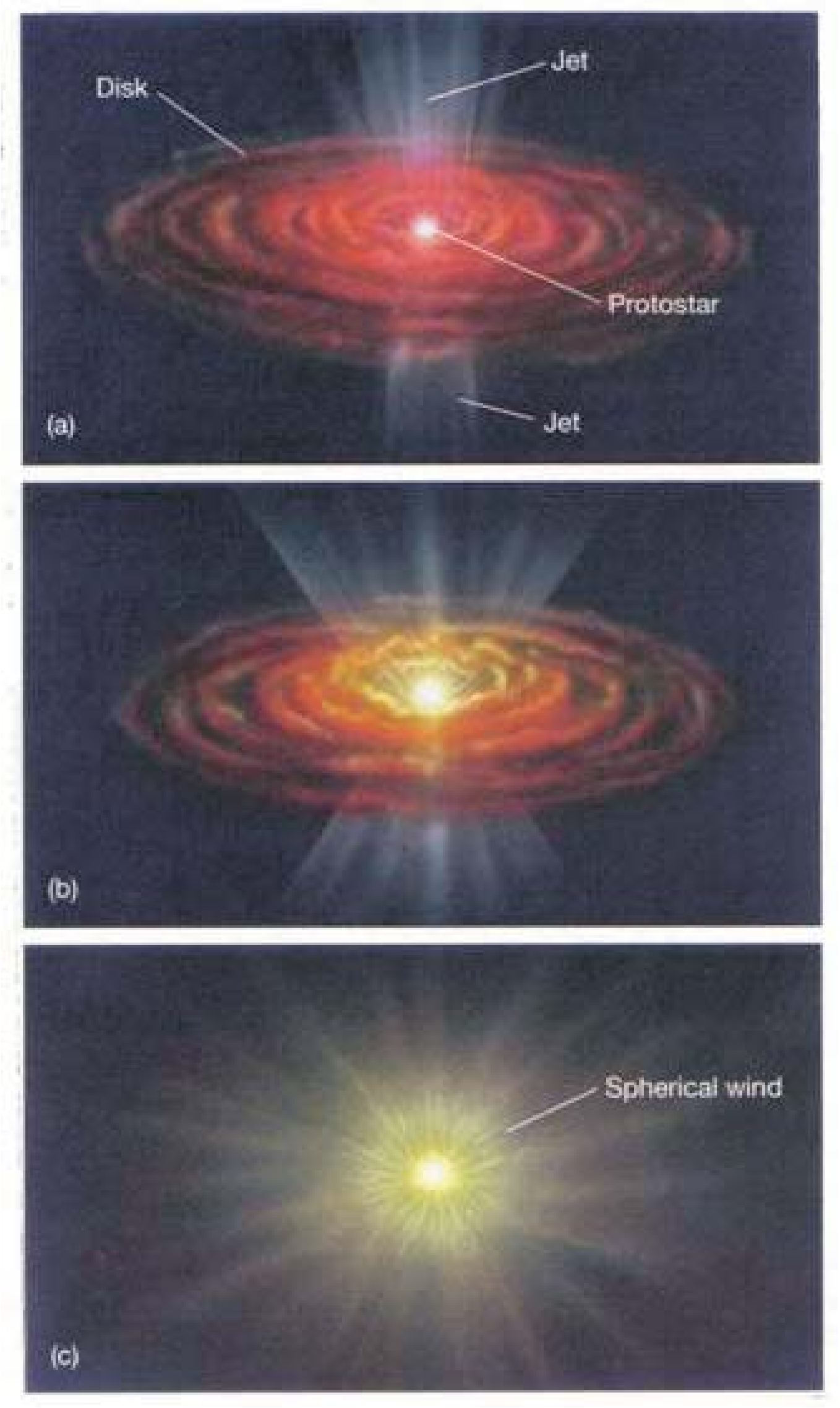}\label{fig4}
  \caption{Stages in the formation of a massive star through accretion in a dense molecular core.}
\end{figure}

In very young SFR, the gravitational contraction of dense molecular cores can give rise to a very rich phenomenology. Molecular clouds are clumpy and highly inhomogeneous. In over-dense regions massive protostars are formed and ultra-compact regions of ionized gas surrounds them. The size of these regions is very small, $\leq 0.05$ pc, implying short lifetimes of the order of $10^{3-4}$ yr. The cloud around the protostar is heated and radiates in the IR with extremely high luminosities that can reach $10^5$ $L_{\odot}$ in some cases. The prestellar core is likely to have angular momentum resulting in the formation of protostellar disks. Evidence for the existence of such disks in been found through IR, optical and line observations (see, eg., Fig. \ref{fig3a}). Magnetic fields, which are thought to be strong in the inner cores of the molecular clouds \cite{Crutcher1999}, are dragged in and twisted by the infalling matter. The result is a magnetic tower that can launch and collimate a thermal jet, with terminal velocities of up to a few thousand km s$^{-1}$. Such outflows are observed in numerous clouds (see Fig. \ref{fig3b}). The accretion rate at this phase can be as high as several times $10^{-3}$ $M_{\odot}$ yr$^{-1}$. The whole formation process, which is illustrated in Fig. \ref{fig4} can last $\sim 10^{5}$ yr.   

As time goes by, the star starts to burn hydrogen and the subsequent wind blows up the circumstellar disk and the jet is quenched. Maser emission is a good tracer of the different aspects of the formation process. Three molecules are of particular importance: H$_{2}$O, OH, and CH$_{3}$OH. This maser emission not only acts as a signspot of active SFRs, but also provides unique information on the physical conditions in the inner regions of the molecular core. 

Water maser emission is usually associated with the presence of molecular outflows and disks. The lines present a wide spread in velocities, that indicating the motions of the molecular material in the region. Hydroxyl masers are indicators of the presence of ultra-compact HII regions. They are always associated with the densest parts of the cloud, being originated in extremely small regions of less than $10^{16}$ cm. Finally, methanol lines arise in shocked gas that is located at the interface between the mass outflow and the external medium. Altogether, these maser lines can shed light on many aspects of the massive star forming process \cite{Garay1999}.

\begin{figure}
  \includegraphics[height=.35\textheight]{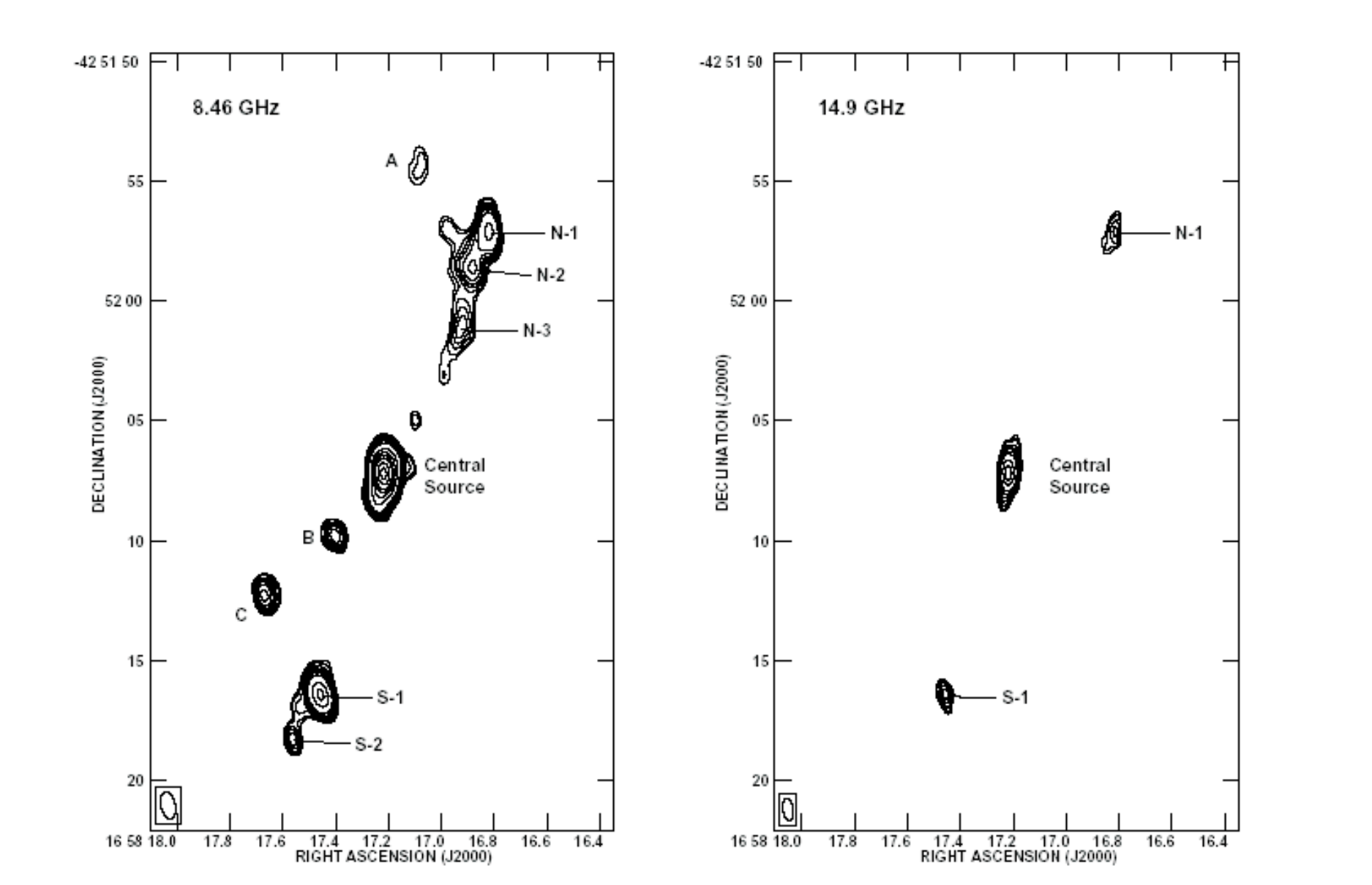}\label{fig5}
  \caption{Radio emission (VLA) from the massive young stellar object IRAS 16547-4247 \cite{Rodriguez}.}
\end{figure}

The jets of massive young stellar objects propagate through the molecular cloud and can break out from its boundaries or can end inside it, remaining hidden at optical wavelength. Radio observations, however, reveal the presence of jets as thermal, elongated features (e.g. \cite{Marti1995}). At the end points of the jets strong shocks are expected to form. These shocks can be sites of diffusive shock acceleration, as indicated by the detection of non-thermal emission of synchrotron origin in a few sources, like IRAS~16547$-$4247 \cite{Garay2003,Rodriguez}, Serpens \cite{Rodriguez1989}, W3(OH) \cite{Wilner} and HH 80-81 \cite{Marti1995}. The first case is particularly interesting, since it seems to be a massive protostar completely embedded in a dense molecular cloud with an extreme IR luminosity of $L \sim 6.2\times 10^{4} L_{\odot}\approx 2.4 \times 10^{38}$~erg~s$^{-1}$. The source is located at 2.9~kpc \cite{Garay2003}. The total
mass of the cloud is $M_{\rm cl}=9\times 10^{2}$~$M_{\odot}$. Molecular line observations indicate that the size of the
cloud is $\sim0.38$~pc in diameter ($\approx 1.1\times
10^{18}$~cm). Assuming a spherical geometry, the averaged particle
(atoms of H) density of the cloud is $n_{\rm cl}\approx5.2 \times
10^{5}$~cm$^{-3}$. The energy density of IR photons in the cloud,
assuming an homogeneous distribution, is $w_{\rm ph}\approx 1.8 \times
10^{-9}$~erg~cm$^{-3}$.

\begin{figure}
  \includegraphics[height=.3\textheight, trim = 0 0 390 550]{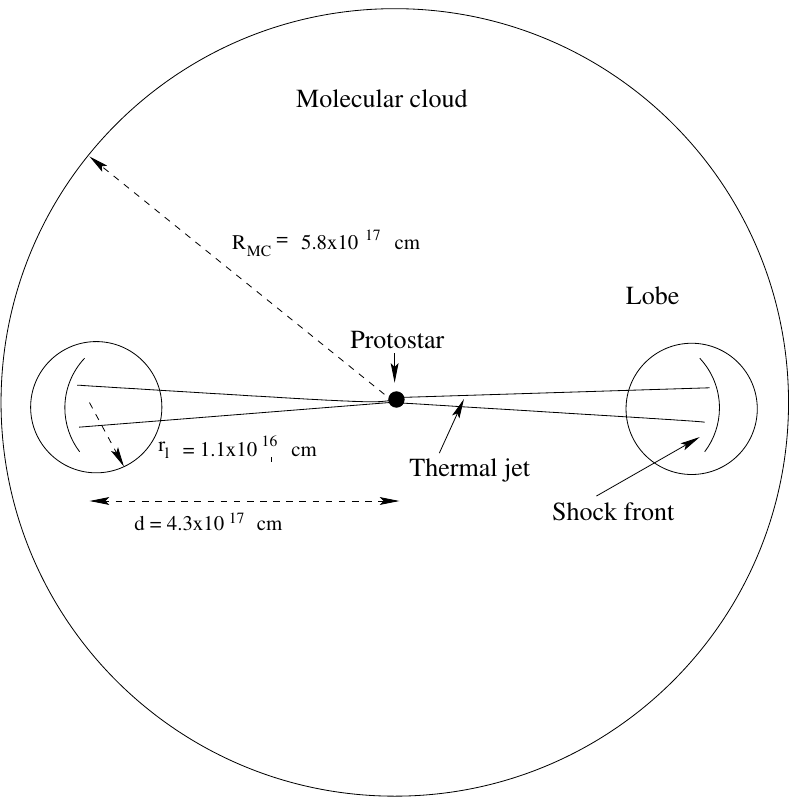}\label{fig6}
  \caption{Sketch of a massive protostar and its associated outflows inside a molecular cloud \cite{Araudo2007}.}
\end{figure}

\begin{figure}[!h]
  \includegraphics[height=.35\textwidth, trim = 50 50 50 10]{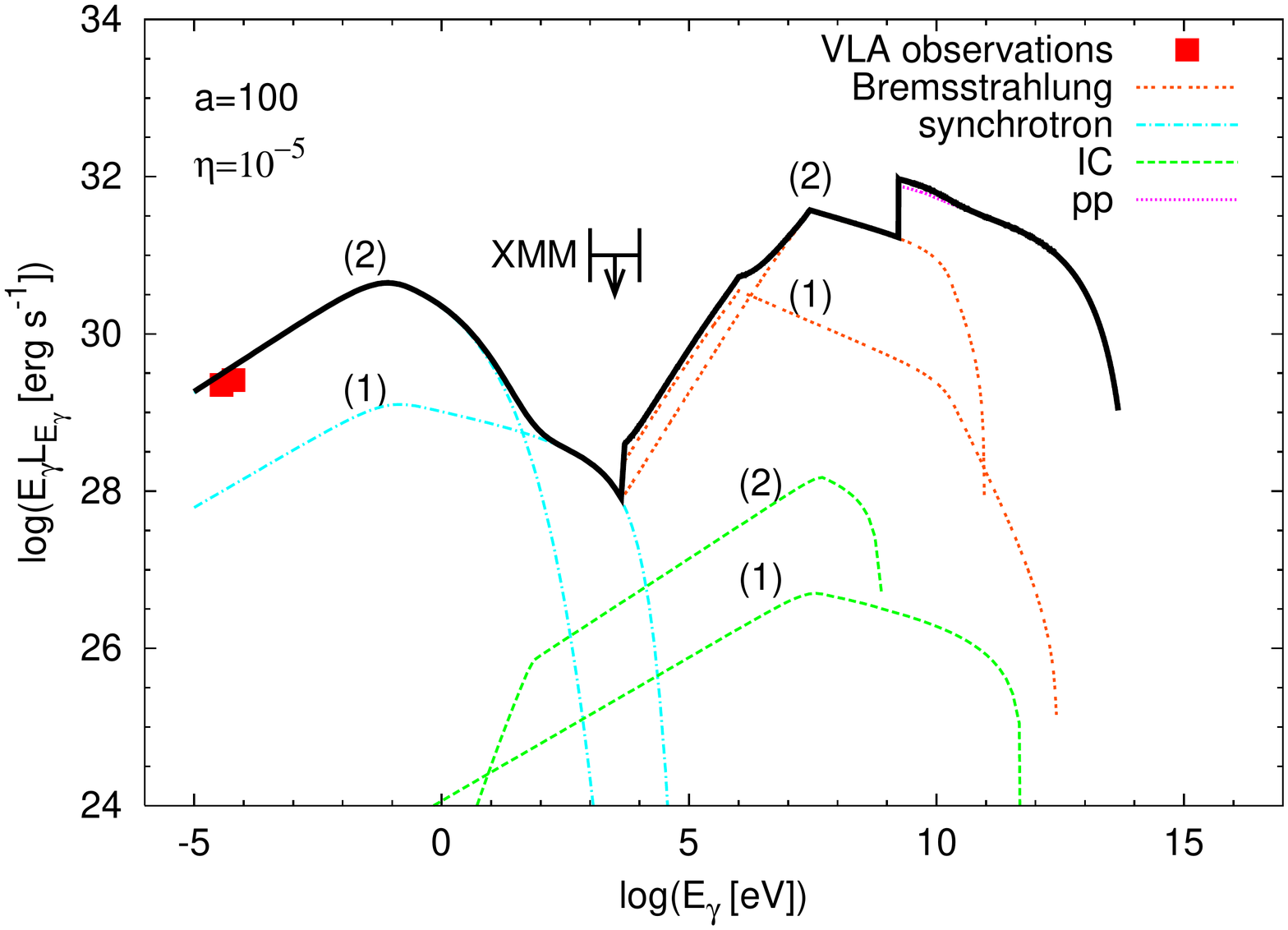}\label{fig7}
  \caption{Spectral energy distribution of the termination point of the jet of a massive young stellar object \cite{Araudo2007}.}
\end{figure}

Radio observations made with the Australia Telescope Compact Array and
the deeper observations with the Very Large Array
show the existence of a triple radio source inside the molecular
cloud (see Fig. \ref{fig5}). The three components of the radio source are aligned in the
northwest-southeast direction, with the outer lobes separated from the
core by a projected distance of 0.14~pc. The central source is
elongated and has a spectral index of $0.33\pm0.05$ ($S_{\nu}\propto\nu^{\alpha}$), consistent with
free-free emission from a collimated jet \cite{Rodriguez}. The integrated
emission from the northern lobe has a spectral index of $-0.32\pm
0.29$, of dubious thermal/non-thermal nature.  The southern lobe,
instead, is clearly non-thermal, with an index $\alpha=-0.59\pm
0.15$. The inferred linear size for this lobe is $\approx 10^{16}$~cm. The whole situation is depicted in Fig. \ref{fig6}.

The presence of relativistic particles in such a complex medium suggest that high-energy emission is likely to occur due to relativistic Bremsstrahlung and inverse Compton interaction with IR photons in th case of electrons, and through inelastic $pp$ collisions in the case of nuclei. The magnetic field in the region of non-thermal radio emission can be estimated assuming equipartition and the obtained values are in agreement with direct measurements by Zeeman effect \cite{Crutcher1999}. Typical fields are of the order of $10^{3}$ G. The acceleration efficiency through diffusive shock acceleration at the terminal jet region in the jet material is small ($\sim10^{-5}$) but enough to yield maximum energies for primary electrons and protons of the order of 3 TeV and 60 TeV, respectively, in the case of a proton to electron density ratio of 100, similar to what is observed in the galactic cosmic ray spectrum. The spectral energy distribution produced by these particles through the different cooling radiative mechanisms is shown in Fig. \ref{fig7}. In this spectrum not only the contribution from primary particles is shown, but also the emission from secondary pairs resulting from charged pion decay chains \cite{Araudo2007}.

From Fig. \ref{fig7} it can be seen that integrated luminosities up to $10^{33}$ erg s$^{-1}$ can result from massive protostars with strong outflows in a SFR. Such emission can be detected with GLAST or with the next generation of Cherenkov imaging telescope arrays, opening a new window to the study of star formation processes and the environment of massive your stars. The combined effect of many protostars in the core of a very young OB association might be even responsible for some unidentified EGRET sources found in the plane in positional coincidence with such associations \cite{Romero1999}. The higher angular resolution of GLAST can differentiate the different individual contributions.  

\section{Conclusions}

Star-forming regions can generate gamma-rays through different processes during their evolution from dark clouds to open clusters. In the early stages of star formation, the contribution from individual protostars can reach integrated luminosities above 100 MeV of around $10^{33}$ erg s$^{-1}$. The composed emission of several of such YSO can illuminate the massive dark cloud in gamma-rays producing a detectable source, even for instruments like EGRET and AGILE. Higher resolution telescopes like LAT of the GLAST-Fermi satellite might resolve the individual sources, then opening a new observational window to the study of massive star formation. Strong IR luminosities, non-thermal radio emission and maser activity are the best tracers for such young protostellar objects.

Once the SFR has aged and early-type massive stars are formed, colliding wind binaries can be significant gamma-ray sources for GLAST-Fermi satellite or even for AGILE. Luminosities up to $10^{34}$ erg s$^{-1}$ at $E>100$ MeV can be expected for WR-OB systems like WR 140. The best indication of efficient particle acceleration in such objects is the clear detection of no-thermal radio emission from the colliding wind region. Protons might escape the system illuminating nearby molecular clouds.

When the SFR has evolved into an open cluster, the collective effect of stellar winds of the strongest O stars and supernova remnants can produce large-scale effects, including particle acceleration up to TeV energies. Such heterogeneous sources can be detected in gamma rays with Cherenkov telescope arrays as extended sources positionally coincident with OB associations and open clusters.      

Regarding isolated massive stars, it seems very dubious to the author whether they can produce a significant amount of high-energy emission. Particle acceleration in the inner stellar winds could hardly be efficient, and in any case losses, both radiative and adiabatic, make very difficult for the particles to reach high-energies. 

Summing up, SFR are promising sites for the production of gamma-rays, by different mechanisms, along their whole evolution. In the next few month and years, star formation studies might incorporate gamma-ray astronomy to probe the most extreme consequences of the formation and evolution of massive stars.    


\begin{theacknowledgments}
I thank Prof. Felix Aharonian for his kind invitation to visit his Group in Heidelberg. I am also grateful to A.T. Araudo, P. Benaglia, V. Bosch-Ramon, and Stan Owocki for fruitful discussions on massive stars and their winds. My work on high-energy astrophysics is supported by CONICET (PIP 5375) and the
Argentine agency ANPCyT through Grant PICT 03-13291 BID 1728/OC-AC. Additional support comes from the Ministerio de Educaci\'on y Ciencia (Spain) under grant AYA 2007-68034-C03-01, FEDER funds.

\end{theacknowledgments}



\bibliographystyle{aipprocl} 


\IfFileExists{\jobname.bbl}{}
 {\typeout{}
  \typeout{******************************************}
  \typeout{** Please run "bibtex \jobname" to optain}
  \typeout{** the bibliography and then re-run LaTeX}
  \typeout{** twice to fix the references!}
  \typeout{******************************************}
  \typeout{}
 }

\end{document}